# Optoelectronic properties of one-dimensional molecular chains simulated by a tight-binding model

Qiuyuan Chen[1, 2], Jiawei Chang[1, 2], Lin Ma[1, 2], Chenghan Li[1, 2], Liangfei Duan[1], Xiaolin Ji[1], Jin Zhang[4], Wei Wu[2],* and Hai Wang[1]*

[1]Key Laboratory of Yunnan Provincial Higher Education Institutions for Organic Optoelectronic Materials and Devices, Kunming University, People's Republic of China.

[2]School of Physics and technology, Kunming University University, Kunming 650091, People's Republic of China

[3]UCL Department of Physics and Astronomy and London Centre for Nanotechnology, University College London, Gower Street, London WC1E 6BT, United Kingdom.

[4]School of Physics and Astronomy, Yunnan University, Kunming 650091, People's Republic of China

wei.wu@ ucl.ac.uk
wanghai_lab@sina.com

## ABSTRACT

Studying optical properties of organic materials is important due to the rapid development of organic light-emitting diodes, solar cells, and photon detectors. Here for the first time we have performed tight-binding calculations for singlet excitons, in combination with first-principles calculations of the excited states in molecular dimers, to describe the optical properties of a zinc-phthalocyanine one-dimensional molecular chain. We have included the intra-molecule and charge-transfer excitations and the coupling between them. Our calculations have successfully interpreted a body of experimental UV-Vis optical spectra of transition-metal phthalocyanines. Compared with the previous ab initio calculations for a molecular dimer, the optical absorptions at the split peaks of the Q-bands can be comparable, which indicates the importance of the coupling between the intra-molecular and charge-transfer excitons.

Successful commercial applications of organic light-emitting diodes (OLED) in mobile phones, tablet and TV, and rapid developments of organic solar cells (OSC) have stimulated an unprecedented volume of experimental and theoretical studies for the optoelectronic properties of organic materials [1, 2]. From the fundamental perspective, the optical properties of organic materials can provide crucial information about the electronic structure and vibrational modes, indicating intrinsic electrical and structural properties and identifying potential materials for optoelectronic devices. The measurement of UV-Vis spectra is a general experimental spectroscopic methodology for analyzing the electronic structure and assessing the potential of organic materials for devices. The comparison between experiments and theoretical modelling is important for understanding the electronic-structure origin of the optical spectra, assessment of the potential of the organic materials for optical applications, and the development of new materials with advanced optical properties. We can elucidate the atomic-scale origin of the optical spectra from modeling electronic structure, providing solid bases for rational molecular tailoring for new materials with advanced optical properties. It is well known that molecular chains in organic crystals are self-assembled exclusively because the strong $\pi-\pi$ electrostatic interactions for the stacking of planar molecules lead to one-dimensional (1D) chain [3, 4]. The primary characteristics of the optical spectra of organic semiconductors such as transition-metal phthalocyanines (TMPc), can be approximated by one-dimensional chains due to the weak electronic coupling between molecular chains. The main features of the optical properties of TMPc, including the Q-band and B-band, originate from the one-dimensional chains in the compound. Serri, et. al have shown that the magnetic properties of cobalt-phthalocyanines (CoPc) are strongly dependent on the 1D chain structure, which can eventually be attributed to the electronic structure of the CoPc single molecule [5]. Similarly, we could therefore understand most of the optical properties of the TMPc organic semiconductors based on the calculations of the optical properties of one-dimensional (1D) molecular chains.

The optoelectronic properties and crystal structures of TMPcs have been studied experimentally extensively. Two UV-Vis absorption bands of TMPcs, namely the B-band and Q-band respectively, have been observed. The main peak of the B-band is located at a wavelength of $\lambda \sim 350$ nm while the main peak of the Q-band at $\lambda \sim 650$ nm [7].

Usually the Q-band of most TMPcs is very broad, splitting into two peaks, for which the interpretation is controversial. Nevertheless, the splitting of the Q-band should originate from the inter-molecular interactions. Although it is generally accepted that the splitting is due to the Davidoff splitting, however the effect of Davidoff splitting is too small if compared with the experimentally observed in the Q-band, which is in the order of 0.1 eV. On the other hand, the observed splitting in the Q-band can also be attributed to the intra-molecular (IM) and charge-transfer (CT) singlet excitations [8, 9].

The optical properties of molecules or finite size nano-structure can be calculated by employing methods such as configuration interaction (CI) [10], time-dependent Hartree-Fock (TDHF) [11], time-dependent density-functional theory (TDDFT) [11], and GW-BSE [12]. The computational cost of configuration interaction (CI) is prohibitively high because the computational cost increases with the number of atoms and basis set factorially. On the other hand, TDHF and TDDFT have been demonstrated as efficient tools for computing the excited states of the molecules or molecule clusters with much lower computational costs. Moreover, the calculation results of TDHF and TDDFT are reasonably good as compared with CI. In practical applications, TDDFT is in general more accurate than TDHF owing to the better description of electron correlations although some of the exchange-correlation functionals is derived from empirical fitting to the experimental observations such as B3LYP [13]. For the excited state of solids, relevant theoretical methods are still in development. For example, Bernasconi, et. al. have tried to extend TDDFT to a periodic solid-state structures in CRYSTAL code [14]. Chan and his colleagues have been developing coupled-cluster methods to compute the excited states in solids [15]. To the authors' best knowledge, it is very difficult [14, 15] to simulate the optical properties of periodic solids from first principles owing to (i) a large number of atoms in the unit cell and (ii) a large number of single-particle excited states need to be taken into account.

Recently TDDFT calculations have been performed for a TMPc dimer, which have shown good accuracy for the excitation energies, but the oscillator strengths for the low energy excitation in the Q-band are too weak to be comparable with experimental observations [16]. However, this low energy excitations at the Q-band are vital for organics-based optical devices, we have therefore established a new theoretical model to improve the description of the oscillator strengths at the Q-band for a molecular chains.

Here we propose an excitonic tight-binding model for a ZnPc 1D periodic chain, in combination with the TDDFT calculations of ZnPc dimer, accounting for the intra-molecule and charge-transfer excitations and the coupling between them. This model strategy is not only suitable for molecular chains, but also any excitonic systems in which a two-level system can be well separated from the higher excited states. The optical spectra of molecule-chain compound TMPc in general have been qualitatively interpreted; we can see the strengthening of the lower-energy band in the optical spectra when including the coupling between excitons. The rest of this paper falls into three parts. The first-principles calculations of a molecular dimer and tight binding model are addressed. The results will be analyzed and discussed, in which the optical spectra of TMPcs are qualitatively interpreted using the model of molecular chain. At the end some general conclusions are drawn.

We have performed TDDFT calculations using Gaussian 09 code [18] for a dimer of zinc-phthalocyanine (ZnPc) molecules (a typical non-magnetic molecule). For the ground-state calculations, we have used hybrid-exchange DFT to compute the Kohn-Sham wave functions and the total energies. We adopted the geometry of the dimer by following the chain structure of the molecular crystal. TDDFT has shown a reasonably accurate description of optical transitions for molecules [16]. Furthermore, the hybrid exchange density functional B3LYP has been chosen for the electron-electron correlations. As suggested by the previous calculations, B3LYP functional can give rather accurate optical spectra and exchange parameters [5, 16]. However, the recent calculations have shown that the range-separated exchange-correlation functional such as CAM-B3LYP can provide much better results, especially for the charge-transfer excitations, which can be used in the future calculations [17]. The parameters of the optical excitations, including the excitation energies and oscillator strengths, will then be used in the tight-binding model calculations.

We employed the tight-bind model that takes into account the intra-molecule and the nearest-neighboring charge-transfer excited states, as shown in Fig.1. The corresponding Hamiltonian to be solved here includes six excitons (two IM and four CT excitons) in a unit cell, which reads

$$\begin{aligned}\hat{H} = \sum_n t_2\, a^{+}_{\mathrm{IM}_1,n} a_{\mathrm{CT}_1,n} + t_1 a^{+}_{\mathrm{IM}_1,n} a_{\mathrm{CT}_2,n} + t_1' a^{+}_{\mathrm{IM}_1,n} a_{\mathrm{CT}_3,n-1} \\ + t_2' a^{+}_{\mathrm{IM}_1,n} a_{\mathrm{CT}_4,n-1} + t_1 a^{+}_{\mathrm{IM}_2,n} a_{\mathrm{CT}_1,n} + t_2 a^{+}_{\mathrm{IM}_2,n} a_{\mathrm{CT}_2,n} \\ + t_2 a^{+}_{\mathrm{IM}_2,n} a_{\mathrm{CT}_3,n} + t_1 a^{+}_{\mathrm{IM}_2,n} a_{\mathrm{CT}_4,n} - \frac{d}{4} a^{+}_{\mathrm{IM}_1,n} a_{\mathrm{IM}_1,n} \\ - \frac{d}{4} a^{+}_{\mathrm{IM}_2,n} a_{\mathrm{IM}_2,n} + \frac{d}{4} a^{+}_{\mathrm{CT}_1,n} a_{\mathrm{CT}_1,n} + \frac{d}{4} a^{+}_{\mathrm{CT}_2,n} a_{\mathrm{CT}_2,n} \\ + \frac{d}{4} a^{+}_{\mathrm{CT}_3,n} a_{\mathrm{CT}_3,n} + \frac{d}{4} a^{+}_{\mathrm{CT}_4,n} a_{\mathrm{CT}_4,n} + h.c. \qquad (1)\end{aligned}$$

Here $a_s^{\dagger}$ and $a_s$ are the bosonic creation and annihilation operators for an individual states as we only consider singlet or triplet excitons. $s$ runs through $IM_{1,2}$ and $CT_{1\text{-}4}$.

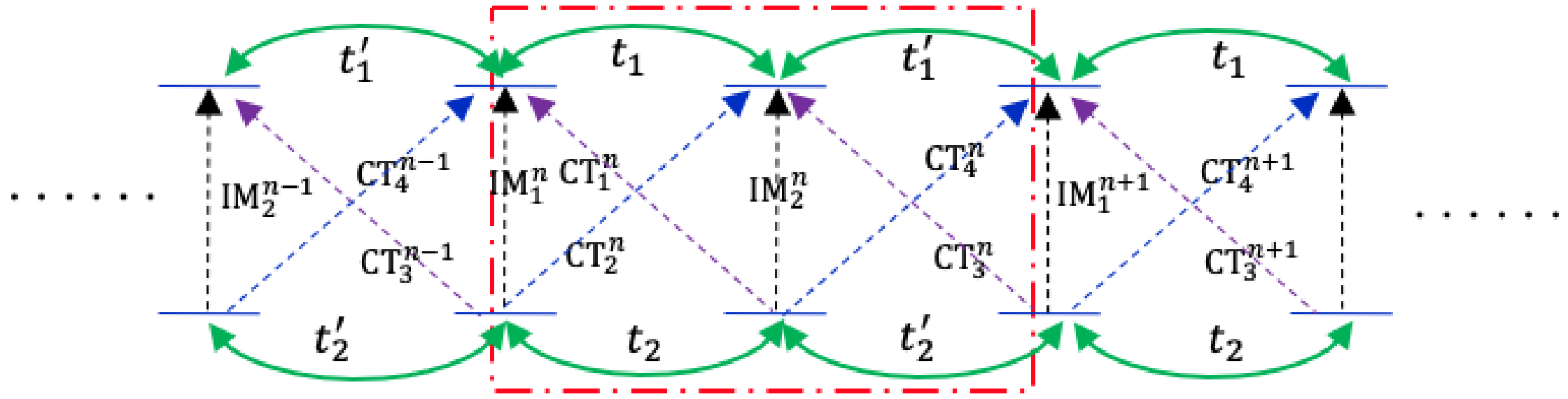

FIG. 1: (Colour online.) The diagram shows the excited states involved in the tight-binding calculations, including intra-molecular ($IM_1$ and $IM_2$) and charge-transfer states ($CT_i$, $i$ = 1, 2, 3, 4). The red box represents the unit cell. We also show the couplings between them by using $t_1$, $t'_1$, $t_2$, $t'_2$, which are between the upper states (normally LUMO) and the lower states (normally HOMO). These can be approximately considered as the couplings between IM and CT excitons.

$t_1$ and $t_2$ are the phenomenological couplings between IM and CT excitons, while the $t'_1$ and $t'_2$ are between cells. For this effective model, the couplings $t_1$, $t_1'$, $t_2$ and $t_2'$ can be variable parameters (real numbers) for the fitting of experimental results. We can also assume that there is a small energy gap $d$ between the intra-molecular and charge-transfer excited states. The Hamiltonian $\hat{H}$ can then be transformed into the $k$-space, $\hat{H}_k$, as follows,

$$\hat{H}_k = \begin{pmatrix} -d/2 & 0 & t_2 & t_1 & t'_1 e^{-ik} & t'_2 e^{-ik} \\ 0 & -d/2 & t_1 & t_2 & t_2 & t_1 \\ t_2 & t_1 & d/2 & 0 & 0 & 0 \\ t_1 & t_2 & 0 & d/2 & 0 & 0 \\ t'_1 e^{ik} & t_2 & 0 & 0 & d/2 & 0 \\ t'_2 e^{ik} & t_1 & 0 & 0 & 0 & d/2 \end{pmatrix}$$

(2)

We can then solve the eigenvalue problems for each *k*-point (assuming the lattice constant here is 1) to obtain the band structure as well as the eigenvectors. We first solve the Hamiltonian to obtain eigenvectors for each *k*-point, and then combined the oscillator strengths for each excited state with the elements of the eigenvector and summed them up in the first Brillouin zone to obtain the oscillator strengths for a molecular chains. Finally, we used Gaussian functions to broaden the lines to obtain the optical spectra of a molecular chains.

We have first computed the electronic structure, especially the excited states of a ZnPc dimer. The Q-band is split into four bands; two of them are CT, and the other two are IM excitations. The CT bands are at ~ 1.6 eV (~750 nm), while the IM bands at ~ 2 eV (~600 nm). The IM bands have a much stronger oscillator strength (~0.5) than CT (~0.02). As shown in Fig.2, we have shown the HOMO (a) and LUMO (b) for a ZnPc dimer, which have predominant contributions to the optical properties at the Q- and B-bands.

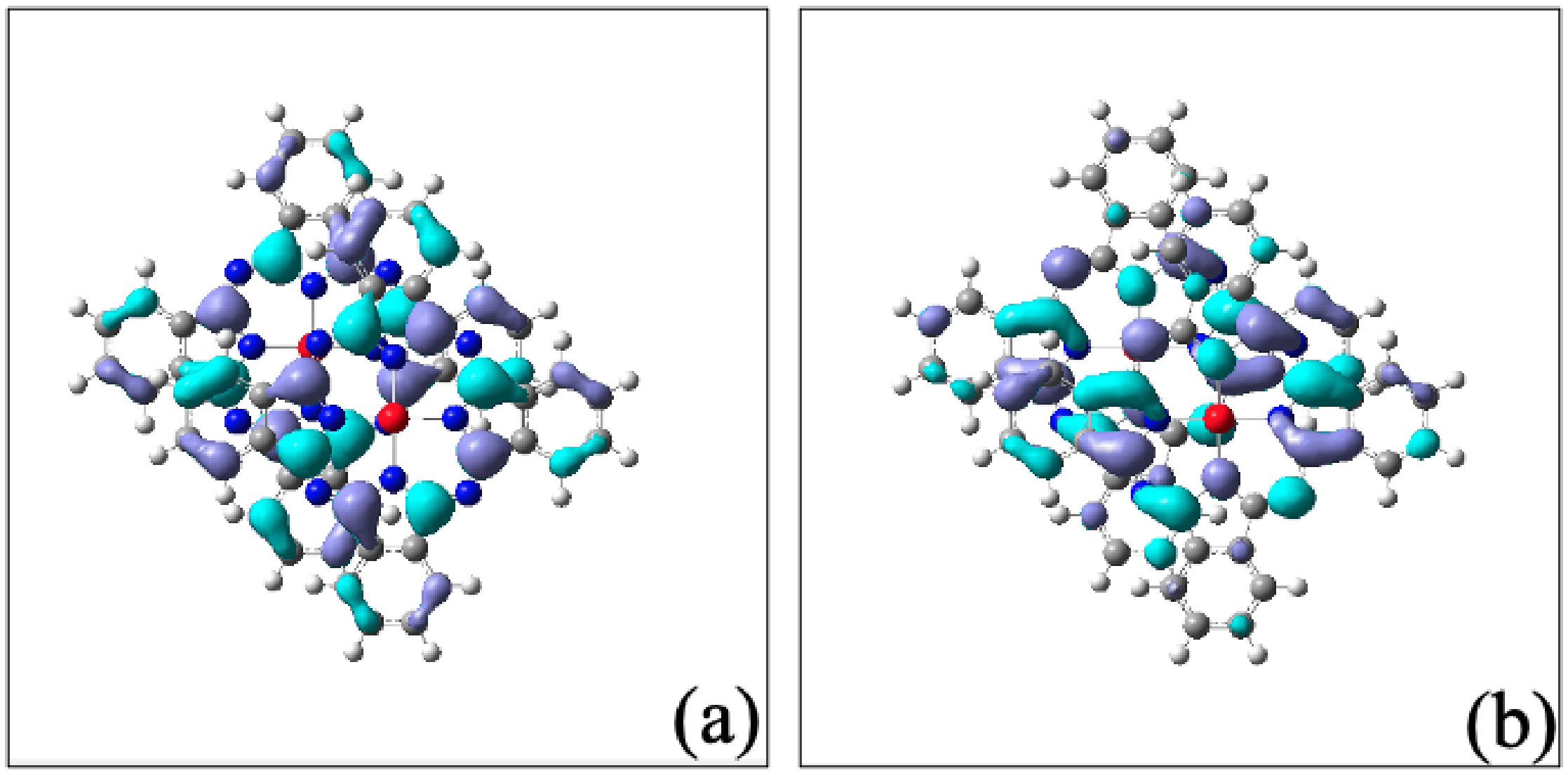

Figure 2: The HOMO (a) and LUMO (b) for a ZnPc dimer are shown. Positive sign is in light purple while negative in cyan for the wavefunction. Zn is red, carbon in gray, nitrogen in blue, and hydrogen in white.

After the first principles calculations, we then solved the eigenvalue problem for the Hamiltonian in eq.2 of a uniform molecular chains, taking $t_1 = t'_1 = 0.05$ eV and $t_2 = t'_2 = 0.005$ eV phenomenologically. The energy difference between the intra-molecular and charge-transfer states is chosen to be $d = \pm 0.025$ eV. The oscillator strengths for the IM and CT excited states are computed as 0.5 and 0.02, respectively, in the Gaussian TDDFT calculations for a dimer. After obtaining the eigenvectors, we subsequently computed the total oscillator strength across the first Brillouin zone from $-\pi$ to $+\pi$ for each *k*-vector. As shown in Fig.3, we have computed the oscillator strength for the above parameters (red and blue curves), and compared this with the calculation in which the coupling is turned off (black curve). As shown in Figure 3, if the coupling is turned off, we only have the strong transition for the IM excited state but negligible CT transition, as expected. When the coupling is turned on, we can observe that the lower-energy excited states are strengthened and become even stronger than the IM excited state when $d$ = 0.025 eV (the blue curve in (a)). The simulated optical spectra are obtained by broadening each transition by using Gaussian functions with a Gaussian width ~0.03 eV. The computed spectra are qualitatively consistent with most of the optical spectra for the transition-metal phthalocyanines [16, 19]. In Figure 3(b), we have shown the experimental UV-Vis spectra for the ZnPc powder (red squares) and nanowires (blue triangles). When comparing Figure 3 (a) and (b), we can see the qualitative agreement between theory (red and blue curves) and experiment. In addition, the computed optical spectra strengthen the long-wavelength part of the Q-band, which improves the previous TDDFT dimer calculations [16]. On the other hand, we can also see the main discrepancy is at ~ 500 nm, where there is a shoulder in the experimental spectra. Moreover, between 700 nm and 800 nm, the experimental spectra are much broader than the theoretical predictions. This could be due to a few aspects, including the addition vibrational modes (not included in the current modelling), other low-lying excitations such as $\pi$–$d$ transitions, and further couplings between molecules along or between the chains. The methodology presented here can therefore be potentially used to fit the experimental optical spectra by varying the parameters including phenomenological couplings, Gaussian broadening, and other fitting parameters. Notice that the energies have been rigidly shifted by ~ 2 eV compared to the above Hamiltonian.

We also computed the excitonic band structure as shown in Figure 2(c) when taking $t_1 = t'_1$ = 0.05 eV and $t_2 = t'_2$ = 0.005 eV, $d$ = 0.025 eV. Here we can think of these transfers integrals as a set of effective parameters that represent the coupling strengths between excitons. We have found these two interesting straight bands that are formed mostly by all the CT states. However, the top and bottom bands contribute most to the mixture between IM and CT excited states. In summary, our calculations can interpret the experimental optical measurements qualitatively. The band structure and corresponding eigenvectors show how the excitonic states can be mixed.

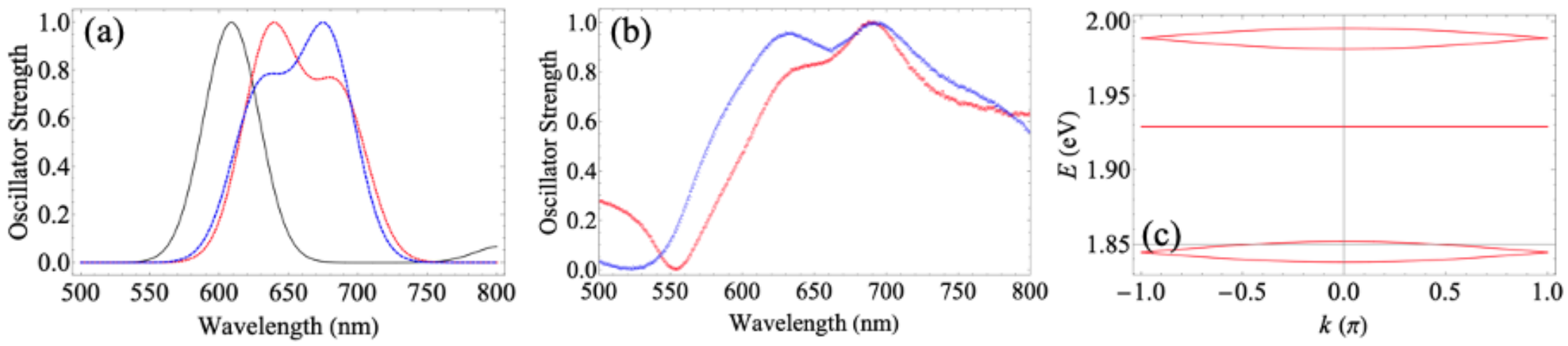

FIG. 3: (Color online.) The computed absorption spectra (a) by using the parameters of Gaussian 09 calculations (red dotted and blue dashed curves) combined with our tight-binding model, which is in a qualitative agreement with the previous experimental results. For the red (blue) curve in (a), $t_1$ = 0.05 eV, $t_2$ = 0.005 eV, $d$ = -0.025 (0.025) eV. In comparison, we have also shown the computed optical spectra with coupling turned off (black curve). For comparison, we have also shown the experimental UV-Vis spectra for the powder (red squares) and nanowire (blue triangles) phases in (b). The corresponding band structure for the blue curve is shown in (c). Especially, we can have a stronger absorption for the longer wavelength when the sign of $d$ is positive.

We can further to generalize our model Hamiltonian to investigate other parameters for inter-cell coupling $t'$ and intra-cell coupling $t$. From here, we adopt $d$ = 0. Besides the uniform chains, we can study similar physics to the SSH model [6], but here for bosons. As shown in Fig.3, we have varied the coupling parameters between excited states in the Hamiltonian and studied the evolution of the band structures when tuning $t_1$ and $t_2$ from (a)-(d) of Fig.3. When tuning $t'_1 = -2.5$ to $t'_1 = -5.0$ and $t'_2 = -1.0$ to $t'_2 = -2.0$ with the same $t_1 = -5.0$ and $t_2 = -2.0$ (from alternating chain to uniform chain), we can observe the band-gap closing between the bands 1 and 2, 5 and 6. Increasing the intercell couplings to $t'_1 = -20.0$ and $t'_2 = -8.0$, the band structure will be gapped once again between the lowest two bands and the top two bands. Notice that the two zero-energy flat band-structure energies (the bands 3 and 4) are not affected by the tuning. They are formed predominantly by the CT states, and these flat bands are robust against variations of dimerization. If including electron-phonon interactions, such as in the SSH model, we anticipate that (i) the model will involve the interaction between exciton and phonon, opening another interesting dimension and (ii) the nontrivial topological properties for photonic band structures could come up if using SSH model, which can be investigated in the future.

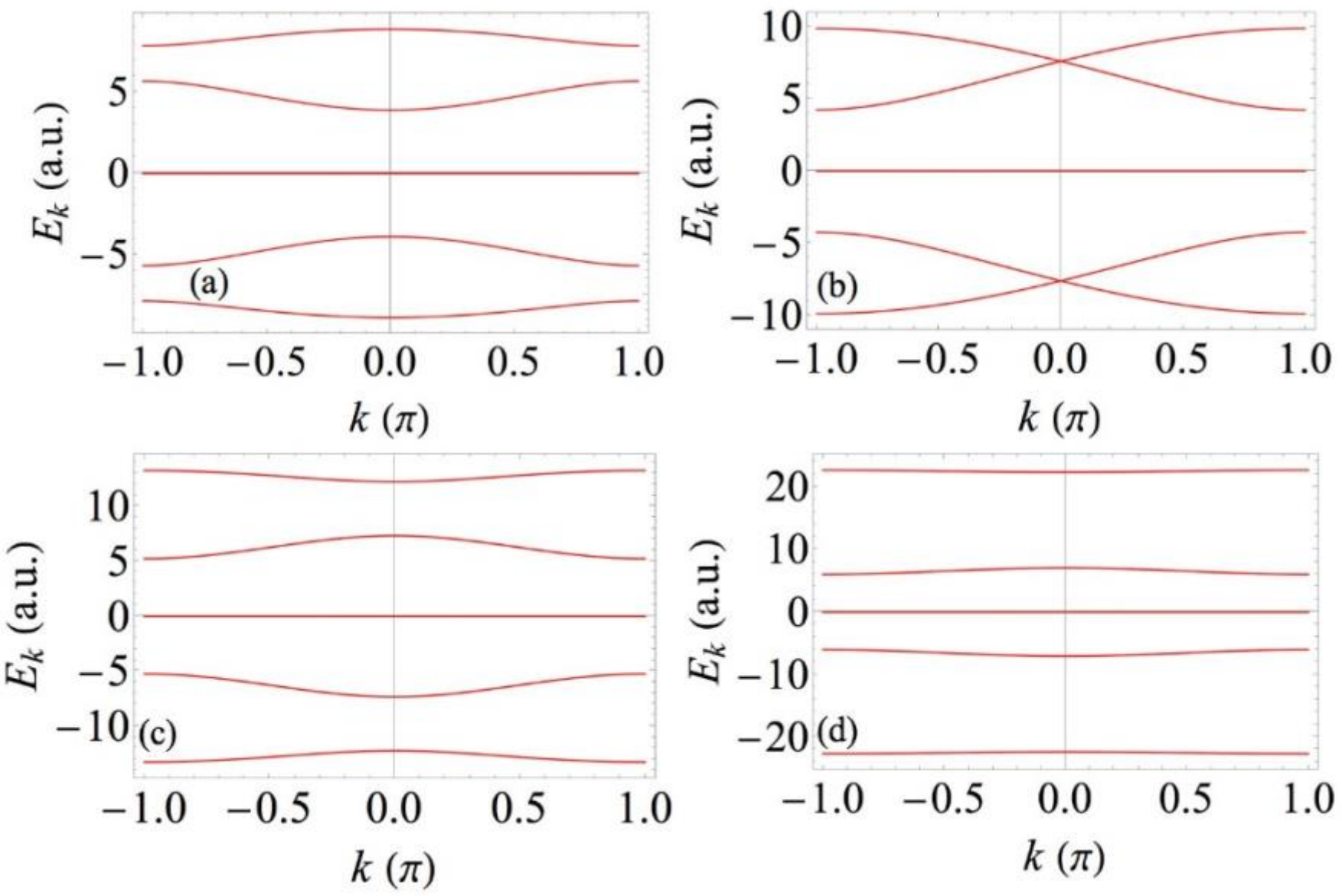

FIG. 4: (Colour online.) The excitonic band structures computed by using our tight-binding model with $d = 0$ are shown. We take following parameters: (a)

$t_1$ =−5.0, $t'_1$ = −2.0, $t_2$ =−2.0, and $t'_2$ = −1.0; (b) $t_1$ =−5.0, $t'_1$ = −5.0, $t_2$ =−2.0, and $t'_2$ = −2.0; (c) $t_1$ =−5.0, $t'_1$= −10.0, $t_2$=−2.0, and $t'_2$ = −4.0; (d) $t_1$ =−5.0, $t'_1$= −20.0, $t_2$=−2.0, and $t'_2$ = −8.0. The zero-energy flat bands are mainly formed by the CT excitons including $CT_1$-$CT_4$.

In summary, we have combined tight-binding modelling of singlet excitons (bosons) and TDDFT calculations to investigate the intra-molecule and charge-transfer excitations and the coupling between them. Our calculations successfully interpreted qualitatively a body of experimental UV-Vis optical spectra of transition-metal phthalocyanines (an organic semiconductor consisting of molecular chains). Our method has low computational cost but can maintain a qualitative feature consistent with the experimental results. Moreover, we can generalize our model Hamiltonian to investigate other parameters for inter-cell coupling *t´* and intra-cell coupling *t* which is analogue to the SSH model. We anticipate there would be more interesting physics related to topological properties.

ACKNOWLEDGMENTS

We gratefully acknowledge Prof. Gong Hao (National University of Singapore) for helpful discussions. We thank the funding from the National Natural Science Foundation of China (Nos. 11564023, 51904137), Yunnan local University Joint Special Funds for Basic Research (2017FH001-007 and 2018FH001-017), the Applied Basic Research Projects of Yunnan Province (2019FD044), Scientific Research Fund of Yunnan Education Department (2019J0563, 2019J0564, and 2020Y0464)， and UK Re- search Councils Basic Technology Programme under grant EP/F041349/1. I thank Bo Chen for stimulating discussions.

DATA AVAILABILITY

The data that support the findings of this study are available from the corresponding author upon reasonable request.

RERFERENCES: